\documentstyle[epsf,rotate]{article}
\begin{document}

\centerline{\bf \Large Sex and recombination in the H\"otzel aging model}
\bigskip

\centerline{A.O. Sousa\footnote{e-mail: sousa@thp.uni-koeln.de \newline After 
01.04.2003, correspondence should be addressed to: ICA1, University of 
Stuttgart Pfaffenwaldring 27, D-70569 Stuttgart, Germany}} 
\bigskip

\noindent{\it Institute for Theoretical Physics, Cologne University, 
D-50937 K\" oln, Germany}
\bigskip

\begin{abstract}
Why sex evolved and it prevails in nature remains one of the great
puzzles of evolution. Most biologists would explain that it promotes
genetic variability, however this explanation suffers from several
difficulties. What advantages might sex confer? The present
communication aims at certain investigations related to this question,
in this way we introduce sexual recombination on the H\"otzel model 
(with males and females) and we compare these results with those from 
asexual reproduction without recombination.
\end{abstract}


\section{Introduction}

Sex, which involves the alternation of meiosis and gamete fusion, is a 
rather inefficient means of self-propagation as compared to asexual 
reproduction, where offspring stem only from a mitotically produced cells. 
One of the most common reasons used to explain the origin and maintenance 
of sex is its ability to reduce the mutation load if consecutive mutations 
lead to an increasing decline in relative fitness, although it is not 
clear {\it a priori} that the heritable variance in fitness is significantly 
increased by sex. Despite decades of developing theoretical models to 
explain why sex is such a widespread phenomenon and how sexual reproduction 
may confer advantages that outweigh its disadvantages, until now no such 
general clear advantage has been found.

Investigations of evolutionary problems by physicists have in fact
boomed in the last few years. Since computer simulations of natural systems
can provide much insight into their fundamental mechanisms, they can
be used to test theoretical ideas that could be otherwise viewed
as too vague to deserve the status of scientific knowledge \cite{book,eigen}.
In this way, many computer models in population dynamics have been 
proposed to investigate the evolution of sex and its justification, as 
well as the comparison between sexual and asexual reproduction, 
for instance, the Redfield model \cite{redfield}, the Penna bit-string 
model \cite{penna}, a genomic bit-string model without aging \cite{tuezel} 
and Stauffer model \cite{dresden,radomski,sousa}.

Of particular interest here is the Heumann-H\"otzel model \cite{hotzel}, 
which originally simulated the evolution of asexual population, composed of 
haploid individuals, without recombination. Thus now we introduce the 
recombination in this model, in order to find out if the sexual reproduction 
(with males and females) can produce better results than the simple asexual 
reproduction \cite{sousa1}. In the next section, we describe the standard 
and the modified Heumann-H\"otzel model, in section 3, we present our 
results and in section 4, our conclusions.

\section{Heumann-H\"otzel Model}

Since it has been proposed by Michael Heumann and Michael H\"otzel in
$1995$, the Heumann-H\"otzel model \cite{hotzel}, which was an unsuccessful
attempt to introduce more ages in the Dasgupta model \cite{dasgupta},
has remained forgotten due to the fact that after many generations it 
reduces to the two-age model of Partridge-Barton \cite{barton}. The 
Dasgupta model consists in taking into account some modifications 
such as hereditary mutations and food restrictions in the
Partridge-Barton model. In fact, the Heumann-H\"otzel paper \cite{hotzel}, 
treats basically the computer 
simulations using the generalized Dasgupta aging model proposed by 
Michael H\"otzel in his work, under the supervision of Dietrich Stauffer, 
in order to obtain the license to teach in German secondary school 
\cite{hotzel2}. Michael Heumann, who was another teacher's candidate, worked 
only on the inclusion of the "Dauer" state in the Dasgupta model 
\cite{heumann}.

Recently, the Heumann-H\"otzel model was reinvestigated and, according to 
the authors, with ``simple and minor change in the original model'' this
incapacity to describe populations with many ages seems to be
surmounted \cite{onody}.

In the original version of the Heumann-H\"otzel, the genome of each
(haploid) individual is represented by one set of probabilities
$p_0,p_1,p_2,...,p_{{\rm maxage}-1}$, where $p_a$ is the survival probability 
that an individual has to reach age $a+1$ from age $a$. At every 
time step $t$, $N(t)*{\rm maxage}$ individuals are chosen randomly to have 
their survival probability $p_a$ altered by mutations to 
$p'_{a}=p_{a}*\exp(\epsilon)$, where the age $a$ is also randomly
chosen. $N(t)$ is the size of the population at time $t$ and
${\rm maxage}$ is the maximum age one individual can live, which is set up
in the beginning of the simulation. The quantity $\epsilon$ is chosen
randomly as any number between $\epsilon_1$ and $\epsilon_2$ and when
it is negative (positive) it corresponds to a deleterious (beneficial)
mutation. 

The effect of food and space restrictions is taken account by an
age-inde\-pen\-dent Verhulst factor, which gives to each individual a
probability $[1-N(t)/N_{\rm max}]$ of staying alive; $N_{\rm max}$ represents
the maximum possible size of the population. This mean-field
probability of death for the computer simulations has the benefit of
limiting the size of population to be dealt with. The passage of time is 
represented by the reading of a new locus in the genome of each individual 
in the population, and the increase of its age by $1$. After taking account 
the natural selection and the action of Verhulst dagger, at the completion 
of each period of life, each individual gives birth to one baby (age=0) 
which inherits its set of probabilities 
($p'_0,p'_1,p'_2,...,p'_{{\rm maxage}-1}$).

In the recent reinvestigation of this model\cite{onody}, individuals 
with age $a$ in the interval $a_{\rm min} \le a \le a_{\rm max}$ 
will generate $b$ offspring and the mutations are allowed only on a 
fraction $F$ ($0 \le F \le 1$) of the babies.

\begin{figure}[!h]
\epsfysize=12,5cm \rotate[r]{\epsfbox{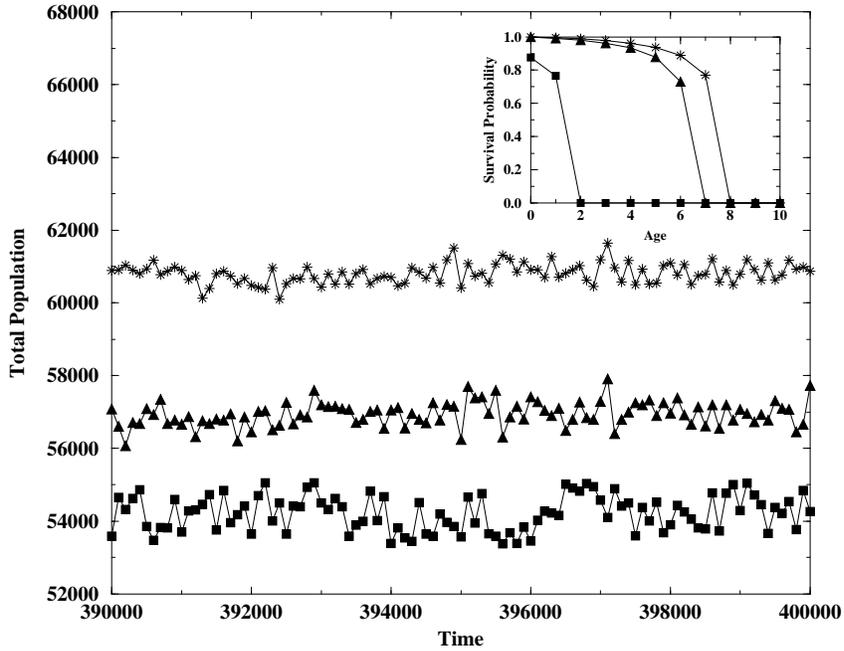}}
\caption{Total population for asexual reproduction (squares), sexual 
reproduction: case (a) (triangles) and case (b) (stars). The inset shows 
the corresponding survival probability.}
\end{figure}

In the sexual version, each (diploid) individual of the population, which
consists of males and females, is genetically represented now by two
sets of survival probabilities, $P(a)^{1}$ and $P(a)^{2}$, to be read in 
parallel. In this way, we have studied the following cases (see below): 
$$P(a)^{1}=(p_0^{1},p_1^{1},p_2^{1},...,p_{{\rm maxage}-1}^{1})$$
$$P(a)^{2}=(p_0^{2},p_1^{2},p_2^{2},...,p_{{\rm maxage}-1}^{2})$$
\noindent
$\bullet$ {\bf Case (a)} - The effective survival probability in some age will 
be the arithmetic average of the values present in both sets at that age:
$$P(a)^{\rm effective}=(\frac{p_0^{1}+p_0^{2}}{2},\frac{p_1^{1}+p_1^{2}}{2},
...,\frac{p_{{\rm maxage}-1}^{1}+p_{{\rm maxage}-1}^{2}}{2})$$

\noindent
$\bullet$ {\bf Case (b)} - The effective survival probability in some age 
will be the maximum value between the values present in both sets at that 
age:
$$P(a)^{\rm effective}=(\max\left[p_0^{1},p_0^{2}\right],
\max\left[p_1^{1},p_1^{2}\right],...,
\max\left[p_{{\rm maxage}-1}^{1},p_{{\rm maxage}-1}^{2}\right])$$

\begin{figure}[!h]
\epsfysize=12,5cm \rotate[r]{\epsfbox{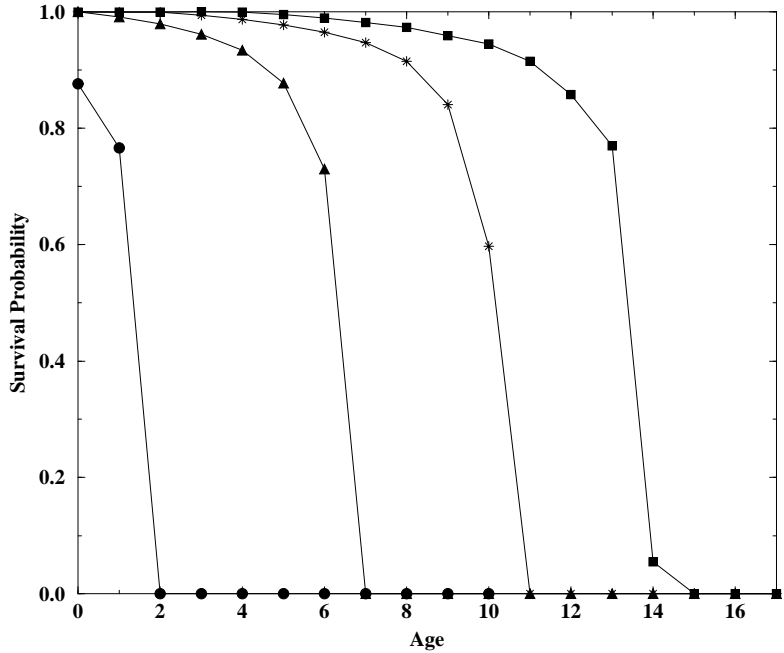}}
\caption{Survival probability for asexual reproduction (circles), sexual 
reproduction with $a_{\max}=17$  and $a_{\min}=1\,{\rm(triangles)}, 
\,3\,{\rm(stars)},\,5\,{\rm (squares)}$.}
\end{figure}

\begin{figure}[!h]
\epsfysize=12,5cm \rotate[r]{\epsfbox{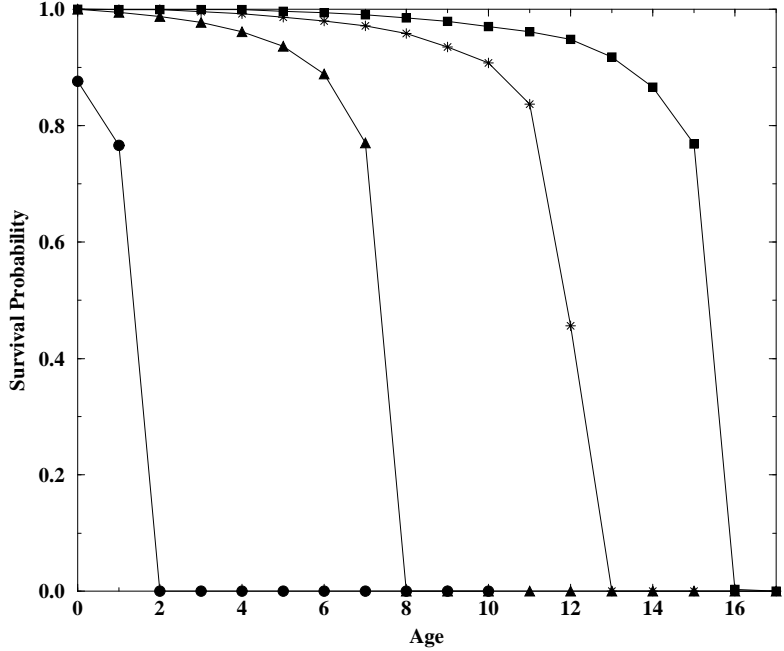}}
\caption{Survival probability for asexual reproduction (circles), sexual 
reproduction with $a_{\max}=17$  and $a_{\min}=1\,{\rm(triangles)}, 
\,3\,{\rm(stars)},\,5\,{\rm (squares)}$.}
\end{figure}

If the female succeeds in surviving until the minimum reproduction age 
$a_{\rm min}$, it chooses, at random, an able male to mate ($a_{\rm min} \le 
{\rm age} \le a_{\rm max}$) and it generates, with probability $p_b$, $b$ 
offspring
every iteration until the maximum age of reproduction $a_{\rm max}$. The
offspring inherits its set of survival probabilities from its parents
in the following way: the two sets of survival probabilities of the
male, for instance, are broken in the same random position, and the
complementary pieces, belonging to different strings, are joined to
form two male gametes. One of the gametes is then randomly chosen to
be passed to the offspring. After that, $m_m$ random mutations are
introduced into this gamete, and the final result corresponds to one
string of the baby genome. The same process occurs with the female
genome, generating the second string of the baby, with $m_f$
mutations. At the end the offspring genome contains a total of
$M=m_m+m_f$ mutations. Finally, the sex of the baby is randomly
chosen, each one with probability $50\%$. This procedure is repeated
for each of the $b$ offspring.

\section{Results}

The simulation starts with $N_{\rm o}$ individuals (half for each sex) 
and runs for $400,000$ time steps, at the end of which (the last $10,000$ 
steps, when the population was stabilized) averages are taken over the 
population. The parameters of the simulations are:

\noindent  $\bullet$ Initial population $N_{\rm o}=10,000$;

\noindent  $\bullet$ Maximum population size $N_{\max}=100*N{\rm o}$;

\noindent  $\bullet$ Probability to give birth ${\rm p_b}=1.0$;

\noindent  $\bullet$ Birth rate ${\rm b}=1.0$;

\noindent  $\bullet$ Mutation rate $m_{m}=m_{f}=1$ per gamete;

\noindent  $\bullet$ $\epsilon_1=0.02$ and $\epsilon_2=-0.04$ (the same 
values used in the original H\"otzel model).
 
\medskip
\noindent Our figures with $N_{\rm o}=10,000$ are confirmed by larger 
simulations with $N_{\rm o}=100,000$, and also by larger simulations 
with $10^7$ time steps.

\begin{figure}[!h]
\epsfysize=12,5cm \rotate[r]{\epsfbox{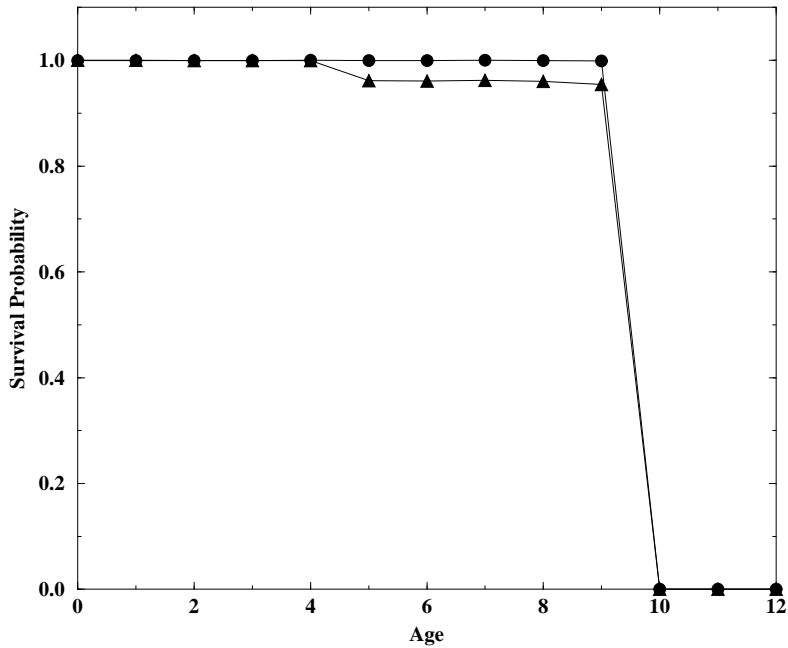}}
\caption{Survival probability for sexual reproduction in case of reproduction 
only at the reproduction age $a_{\max}=a_{\min}=10$ (circles), and in 
case of reproduction at some age between $a_{\min}=5$ and $a_{\rm max}=10$ 
(triangles).}
\end{figure}

From Figure 1 and its inset we can see that the diploid sexual population 
is not only larger than the haploid asexual one, but also presents a higher 
survival probability.

\begin{figure}[!h]
\epsfysize=12,5cm \rotate[r]{\epsfbox{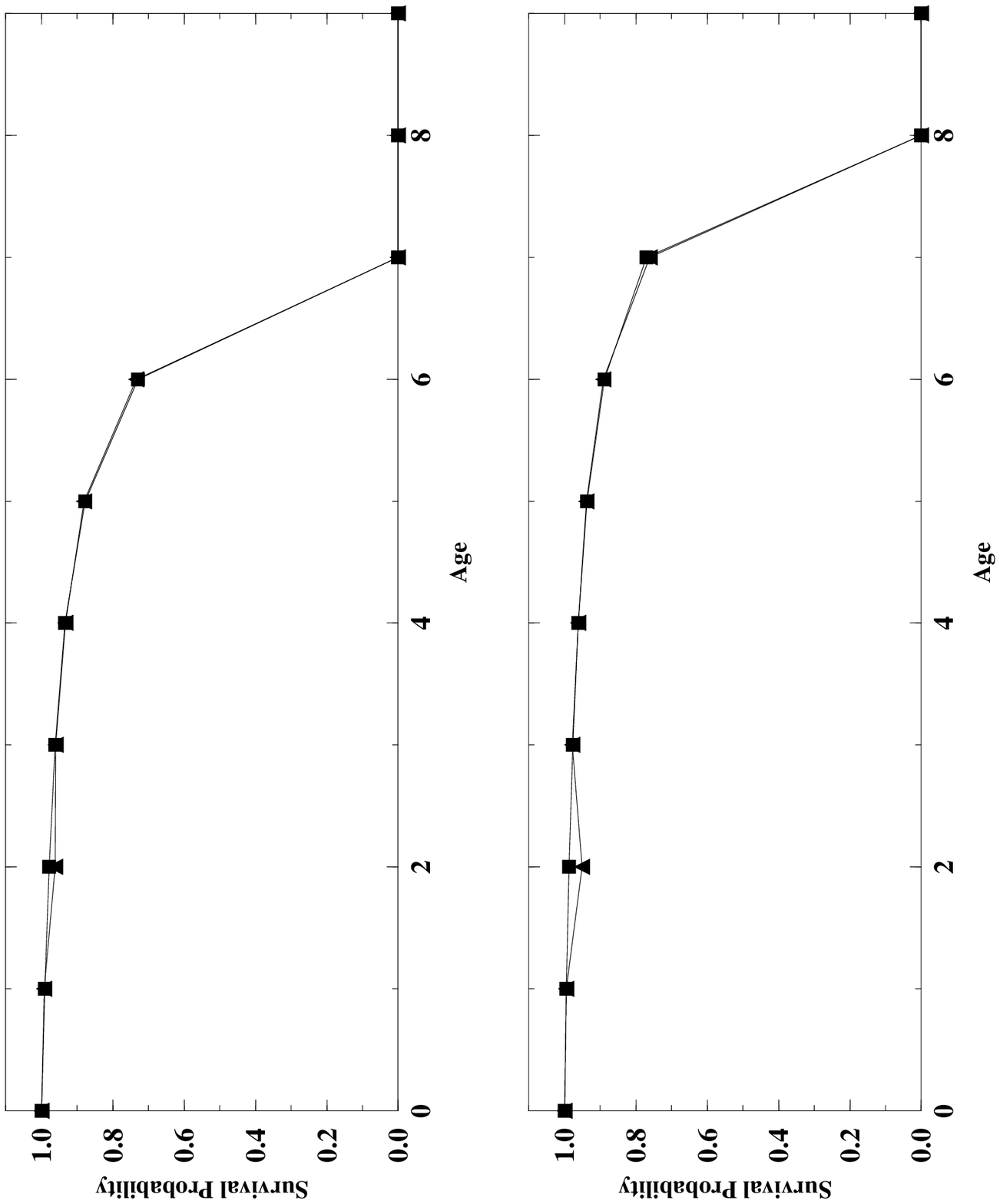}}
\caption{Survival probability for asexual reproduction (circles), sexual 
reproduction with $a_{\min}=1$,\,$a_{\max}=17$  and $d=1\,
{\rm(squares)}, \,2\,{\rm(triangles)}$. Case (a) (top) and case (b) (bottom).}
\end{figure}

\begin{figure}[!h]
\epsfysize=12,5cm \rotate[r]{\epsfbox{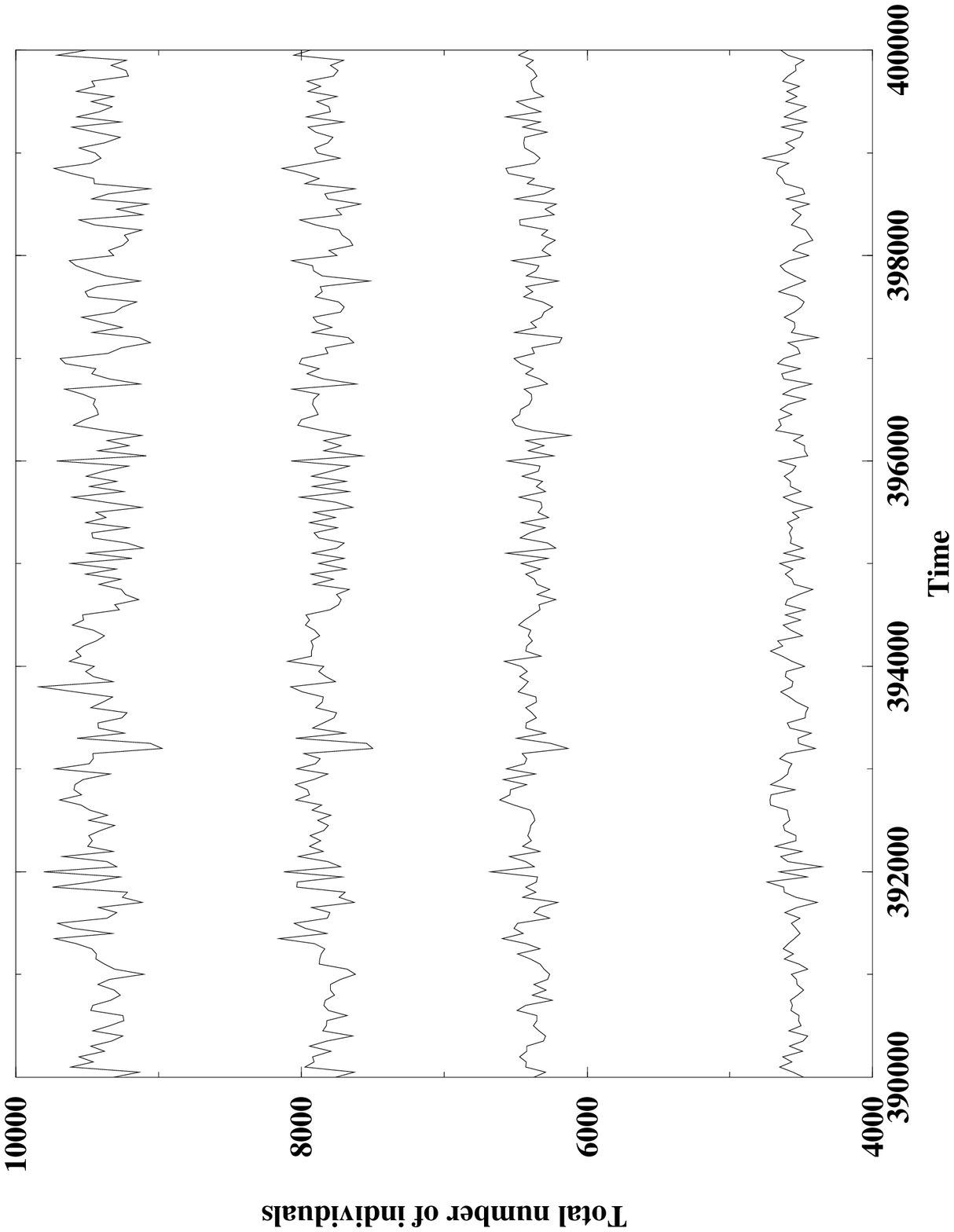}}
\caption{The total population for ages 0, 1, 2, 3 (top to bottom, 
respectively) as a function of time for sexual reproduction when the 
mutations are exclusively harmful ($\epsilon_1=0.0$ and $\epsilon_2=-0.20$).}
\end{figure}

In Figure 2 (case (a)) and Figure 3 (case (b)), we present the survival 
probability as a function of age for different period of reproduction 
($a_{\min} \le i \le a_{\max}$). {\it Aging starts with reproduction}: 
the survival rate decays as soon as reproduction age is reached. There are 
no individuals alive older than the maximum reproduction age $a_{\max}$. 
Figure 4 corresponds to the case in which all the individuals of the 
population reproduces only once - the so-called catastrophic senescence 
effect \cite{barton,salmon}. In this way, two rules of reproducing were 
adopted: 1) The reproduction age is the same for all individuals 
($a_{\max}=a_{\min}=10$), 2) The reproduction age is randomly chosen between 
$a_{\min}=5$ and $a_{\max}=10$. We can noticed that this effect is more 
pronounced for the former \cite{barton,salmon}, as well that the responsible 
for that are both breeding once and breeding for all individuals at the same 
age. The explanation for these effects observed in Figures 2-4 is based on 
the Darwin theory: individuals must stay alive in order to reproduce and 
perpetuate the species. If they can no longer generate offspring but 
remain in the population, they are killed by the accumulation of deleterious 
mutations \cite{book}.

In fact, real mutations can be divided into the common recessive
(almost $90\%$ of the real mutations are recessive) and the rare
dominant mutations. In this way, if among the many genes of a species,
one of the father's genes differs from the corresponding one
of the mother, then it adversely affects the child only if the
mutation is dominant. Recessive mutations affect the child only if
both the father and mother have them. In order to take into account this
aspect in the sexual version of H\"otzel, at the beginning of
the simulation we choose randomly $d$ dominant positions and keep
them fixed during the whole simulation. The effective survival
probability in the dominant positions (ages) will be the smallest value of 
the two located in the same position in both strands, and
for recessives ones the effective survival probability will be the
arithmetic average of them. In Figure 5, we can see that the inclusion of 
dominance does not alter the lifespan of the population, although it has 
been observed that population evolved without dominance is larger than 
the other without dominance due the deaths in the former being bigger than 
the latter, since in these dominant positions the effective survival 
probability is the minimum value between the values present in both 
sets at that age. In the particular simulation shown, for ${\rm age}=2$ it 
is noticed a decrease in the survival probability when the dominance is 
considered, since the ages $2$, $13$, $10$, $15$ were dominant positions.

Figure 6 shows the time evolution of the total population of each age
for sexual reproduction when the mutations are exclusively harmful. From
this figure we can notice that a stable population for ages $a\le3$ is
obtained, in contrast to the original model in which there are no
individuals alive older than age $a>2$, even if beneficial mutation
and also a deleterious mutation rate $5$ times smaller have been
assumed. The result obtained here (Fig. 1), introducing sex in the
original model, was found with the asexual H\"otzel model \cite{onody} only
when mutations were allowed on a very small fraction $F=10\%$ of the babies
and also a minimum age of reproduction $a_{\rm min}=8$ was considered. In 
our simulations, $F=100\%$, $a_{\min}=1$ and $a_{\max}=17$.

\section{Conclusions}

We have shown that main problem related to the H\"otzel model, which was 
its incapacity to treat populations with many age intervals, has been 
overcome by introducing recombination (with males and females) in this 
model, without any other assumptions. As well as, with the inclusion of sex 
in the model, the population meltdown observed in the asexual version, 
when only deleterious mutations are considered, has been avoided. Moreover, 
in agreement with some earlier models, we have also obtained that the 
sexual reproduction (with males and females) produces better results than 
the asexual one. 

\bigskip
\noindent{\bf Acknowledgments}

I would like to thank Suzana Moss and Dietrich Stauffer for discussions 
and a critical reading of the manuscript, and DAAD for financial support.

\end{document}